# The fracture toughness of molybdenum at different grain sizes under and since brittle-ductile transition: computer modeling and experiment

K.M. Borysovska[*], N.M. Marchenko, Yu. Koval, Yu.N. Podrezov, S.A. Firstov

Frantsevich Institute for Problems of Materials Science, 3 Krzhizhanovsky str., 03142 Kyiv, Ukraine

[*]e-mail: *kmborysovska@ukr.net*

*The sharp growth of the fracture toughness after brittle-ductile transition happens at grain sizes approximately equal to the plastic zone size. Here we analyze the influence of the grain boundary on the evolution of the ensemble of dislocations near the crack tip using dislocation dynamics method. We show evidence that for large grain sizes, the size has little effect on the ensemble of dislocations and the fracture toughness, but when dislocations reach the grain boundary, distribution of dislocations changes, which leads to increase of the fracture toughness.*

Keywords*: fracture toughness, grain size, brittle-ductile transition, dislocation ensemble.*

1. **Introduction**

Numerous experiments indicate that the fracture toughness of metals depends on the fracture mechanism and the type of material [1-3]. Many experimental data relate fracture toughness to grain size using a formula similar to the Hall-Petch law: $K_{cr}=K_0+\kappa d^{-m}$ in the case of average homologous temperatures. The values of m are quite different for different classes of materials, usually ranging from 0.5 to 1 [4-9]. However, if in the case of the classical Hall-Petch equation the relationship between grain size and yield strength is strictly substantiated, then in the analysis of fracture toughness such a relationship is not apparent due to the lack of strict physical models linking the interaction of a crack with an ensemble of dislocations. In our previous work [10], performed on a molybdenum polycrystal demonstrating a ductile-brittle transition in the test temperature range, it was shown that the dependence of fracture toughness can be quite complex. If the size of the plastic zone is commensurate with the grain size, a change in the mechanism of formation of the plastic zone and the mechanisms of destruction of materials may occur [1]. These issues require more careful study, both experimental and theoretical.

There are a large number of publications on the ductile-brittle transition performed on single crystals. First of all, these are the works of Roberts et al. performed on silicon, molybdenum and intermetallics TiAl [11]. Different groups of researchers have conducted experiments on single-crystal materials: bcc metals: iron-silicon [12] and molybdenum [13]; ceramics: sapphire [14], germanium [15], magnesium oxide [16] intermetallics: TiAl [11], NiAl [17]. The general conclusion: the temperature dependence of the mechanisms of nucleation and movement of dislocations leads to a change from brittle fracture to ductile.

Therefore, accurate accounting of the number of dislocations and their positions near the crack tip will allow describing the dependence of the crack resistance of metals on temperature. This problem is a problem of many materials bodies, therefore, computer modeling methods, in particular dislocation dynamics, are most suitable for solving this problem. It should be noted that the use of modern 3D models for example, Kubin [18], which describe the three-dimensional processes of nucleation and movement of dislocations, significantly complicated calculations. Therefore, 2D models are often used. In particular, in [19], the behavior of single crystals with edge dislocations is modeled with a numerical solution of the image forces. This work takes into account the resistance of the lattice to dislocation movement, the nucleation of dislocations, their interaction with obstacles and annihilation.

In the case of polycrystals, the analysis is somewhat more complicated. In this case, the most interesting model objects are high-carbon steels [2,4] intermetallics [6,20] or refractory metals tungsten [21,22], molybdenum [10], where the grain size is comparable to the size of the plastic zone under test conditions near room temperature. Under these conditions, a sharp increase in plasticity or crack resistance is observed with increasing temperature or decreasing grain size. Mechanical models of Irwin [23] or the more complex model of Hutchinson, J. W., Rice, J. R., Rosengren, G. F [24], where hardening is taken into account, cannot explain this effect, since they do not take into account dislocation behavior. In the work of Pippan and Rimelmoser [25], the movement of dislocations is modeled. The results of the dislocation model are compared with the predictions of the continuous description of cyclic crack tip plasticity. Calculations showed that both the dislocation model and the mechanics of the continuous medium lead to the same result at high load levels, but they differ significantly at low stress intensities.

These results confirm the feasibility of using dislocation models to analyze the mechanical behavior of polycrystals, taking into account the interaction of dislocations that are simulated from the crack tip with the grain boundary. Pioneering contributions in the areatopic of thermally activated dislocation dynamics were made by Li and Li [26]. They were the first to use the method of dislocation dynamics to study the patterns of their propagation in a polycrystal, taking into account thermal activation processes. However, the modeling unexpectedly showed that grain refinement worsens the fracture toughness. To overcome this contradiction, the work of X.H. Zeng, A. Hartmaier [27] took into account the process of crack blunting, as well as the influence of neighboring grains and loading rate. These refinements improve the results on the influence of grain size on crack resistance.

An improvement of this model was made in the work of Jens Reiser & Alexander Hartmaier [28] in which the crack crosses several grains and interacts with several boundaries, forming Frank-Reed sources at the intersection points. This increases the number of sources with decreasing grain size and, as a result, improves the plasticity of fine-grained materials. The authors obtained good agreement between theory and experiment. This extremely interesting modeling allows us to significantly advance our understanding of the dislocation mechanism of the formation of the ductile-brittle transition formation. However, this model does not consider Frank-Reed sources in neighboring grains.

The influence of Frank-Reed sources in neighboring grains was considered in [29], an increase in crack resistance was obtained, which depends on both the size and transparency of the grain boundaries. However, in this work grains no larger than 5 nm were considered. In our opinion, it is the peculiarities of the interaction of the plastic zone, or rather the dislocation pile-up, from which the Frank-Reed source is triggered, in the neighboring grain, that can be the determining factor for that determines the change in the mechanisms of dislocation motion in the plastic zone and enhances the shielding effect.

The purpose of this work is an experimental study of the effect of grain size on crack resistance in the region of the ductile-brittle transition as well as and modeling of the interaction processes of dislocations simulated by a crack with grain boundaries, on which Frack-Reed sources are located, by the method of 2D dislocation dynamics method.

The paper is organized as follows: the experimental method described in the second section, experimental results and it's discussions in sections 3. Computer model explained and the results of numerical simulations are presented in sections 4 and 5 respectively.

## 2. Materials and methods

The experiments were performed with molybdenum alloy (Zr-0.15%, C-0.003%). The samples were annealed (2 hour) at different temperatures (1673-2223 К), the grain sizes are shown in Table 1.

Table 1. Grain size molybdenum alloy after annealing.

| T, K | 1673 | 1773 | 1873 | 1973 | 2023 | 2073 | 2223 |
|------|------|------|------|------|------|------|------|
| d, µm | 42 | 82 | 165 | 200 | 280 | 400 | 600 |

Rectangular samples were 5 mm tall and 3.5 mm wide, distance between supports was 20 mm. Notching was obtained in the middle of the long dimension by electrospark machine cuts: 1,5 mm deep and 0,05 mm wide. The sharp spark crack was 100 µm at the tip of the notching.

The samples were subject to three point bending at a rate of 0,1 mm/min and temperatures 77 and 293 K. Photometric experiment showed that in all experiments crack started abruptly without stage growing up as in brittle material allowing calculation of fracture's toughness by formula in accordance with the Knott [30]:

$$K_{1c} = \frac{3PL}{BW^{3/2}}\left(1,93\left(\frac{a}{W}\right)^{1/2} - 3,07\left(\frac{a}{W}\right)^{3/2} + 14,53\left(\frac{a}{W}\right)^{5/2} - 25,11\left(\frac{a}{W}\right)^{7/2} + 25,8\left(\frac{a}{W}\right)\right) \quad (1)$$

where P – fracture load, L - length, W - sample width, B – a/2, a - crack length. Observational error is 5%.

### 3. Experimental results and discussion

Our experiments revealed a different behavior of relationship between the fracture toughness and the grain size at different temperatures that is shown in Fig. 1. At 77 K fracture toughness is slowly increasing with the grain size. At room temperature it demonstrates a sharp growth at the grain size of about 280 µm. Such behaviour disagrees with the typical fracture toughness – grain size dependence, which we study further. A similar effect was observed in steels, due to carbide particles [31],[32] and the same effect was found on tungsten [33].

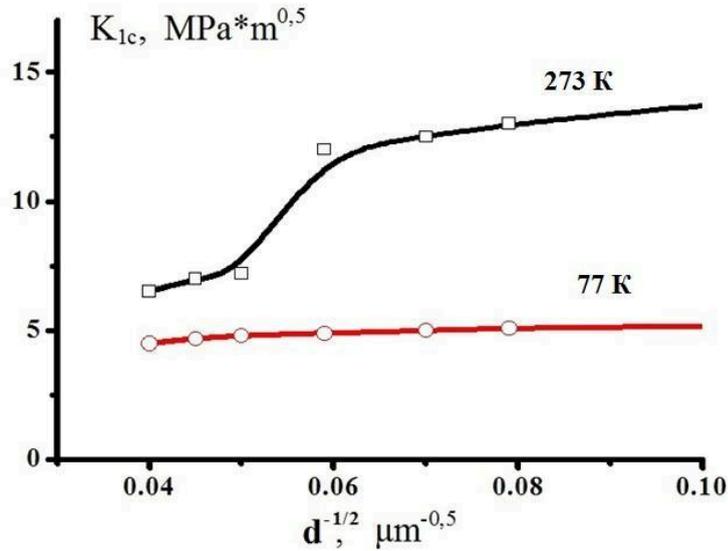

Fig. 1   Dependence of the fracture toughness on the grain size for low-alloyed molybdenum.

According to fracture mechanics the size of plastic zone can be calculated as [34]:

$$r_P = \frac{1}{2\pi}\left(\frac{K_{1c}}{\sigma_T}\right)^2 \qquad (2)$$

where $\sigma_T$ is the yield point.

To estimate the size of the plastic zone near the crack tip we have additionally measured yield point value and investigated its structure sensitivity. Fig. 2. shows the dependence of yield point on the grain size and obeys classic Hall-Petch equation. In Fig. 3 the plastic zone size, calculated according to Eq. 1 is plotted versus the grain size.

The absence of the fracture toughness jump at liquid nitrogen temperatures can be explained based on the assumption that the plastic zone is significantly smaller than the grain size even at very small grain sizes. Unfortunately, it is not possible to measure yield point at 77 K directly, since all samples are destroyed due to brittleness at this temperature. However, it is possible to estimate the yield point of molybdenum at this temperature following the method of Milman [34], where it was obtained by recalculation from the hardness data and its value was estimated about 1000 MPa. Substituting this $K_{1c}$ value into Eq. (1), we obtain $r_p$ of about 3-5 μm.

The dependence in Fig. 3 has S-like shape and its inflection point is located at the grain sizes of about 280 μm, which corresponds to the condition of $d=2r_p$, and therefore the jump of fracture toughness corresponds to the point where the plastic zone size becomes comparable to the grain size.

In fig. 4 one can see a small plastic deformation in front of the crack, which occurred before it reached the Griffet's size. The ensemble of dislocations in front of the open crack in pure molybdenum is shown in fig. 5.

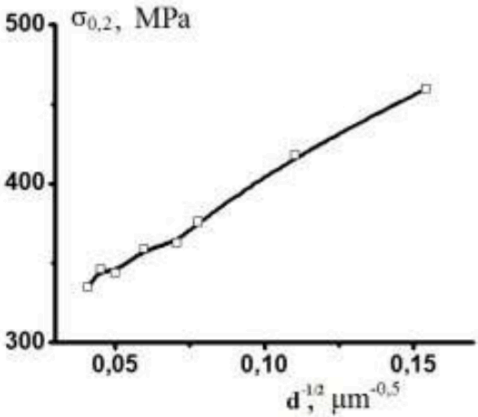
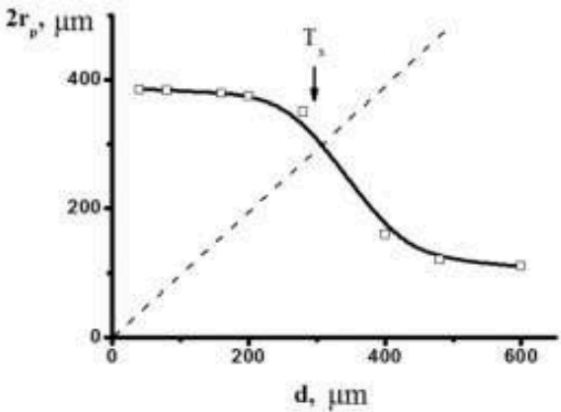

Fig. 2. Dependence of the yield point on the grain size for the low-alloyed molybdenum.

Fig. 3. Dependence of the plastic zone size on the grain size for the low-alloyed molybdenum.

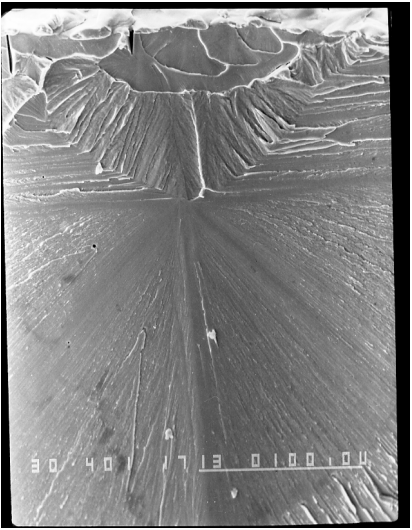
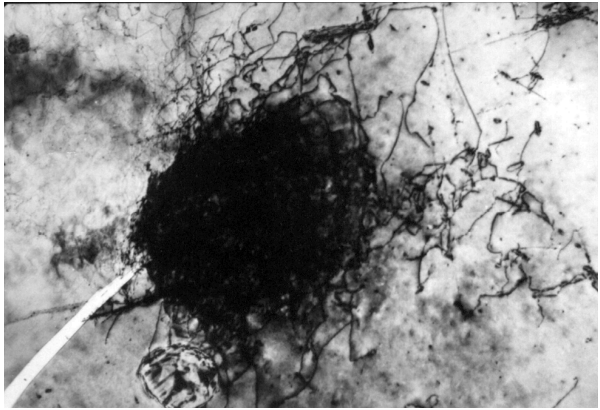

Fig. 4 Fractography of molybdenum alloy.

Fig. 5 Dislocation ensemble near crack tip in pure molybdenum.

Theoretically it is expected from the analysis of the plastic zone size in Eq.(1), that the plastic zone size must grow abruptly. But it is likely that the observed jump in the fracture toughness is caused not just by the sharp increase of the plastic zone size, but rather by the redistribution of the dislocations in the plastic zone due to their interaction with the grain boundary.

## 4. Computer modeling

To understand this process in more detail, we performed modeling of the evolution of the dislocation ensemble near the crack tip under stress using two-dimensional dislocation dynamics (DD) method. Modeling of the dislocation ensemble near the crack tip with this method was done previously [35],[36],[37],[38],[27],[39],[40]. In particular, Zeng et al., 2010 [27] studied the interaction of the grain boundary with the dislocation ensemble near semi-infinite atomically sharp crack and found that the stress intensity factor grows when the plastic zone touches the grain boundary. Similar results were obtained by Noronha and Farkas [36] with molecular and dislocation dynamics methods taking into account crack blunting when dislocations were emitted.

The DD method is a variation of computer modeling of the many-body problem. The time consists of discrete steps. At each step, one calculation of stress is done for all dislocations in the ensemble, then velocities are calculated and the corresponding new positions (Fig. 6). More about this method can be found in [41].

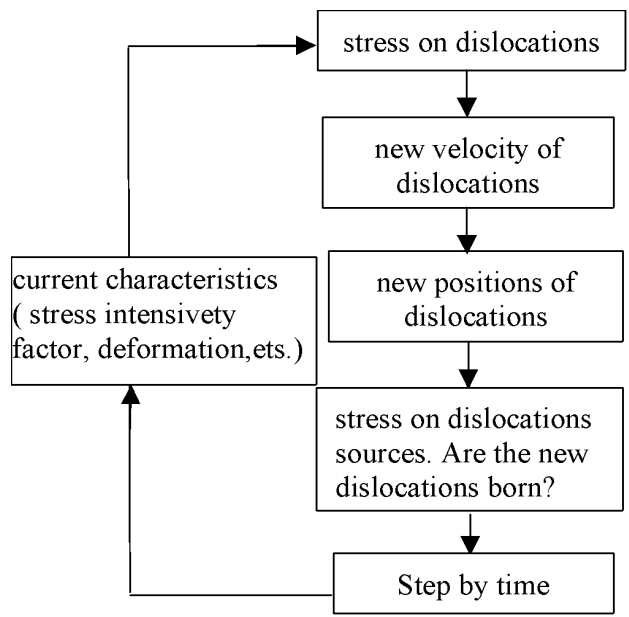

Fig. 6    Flowchart of dislocation dynamics method.

First, the stress on the dislocations was calculated. The stress on every dislocation or dislocation source consists of the following components Ohr [42]: repulsive external stress [43], attractive stress from the crack image [43] and the crystal lattice frictional stress, the influence of friction stress on dislocation ensemble show in [44], [45], [46] and stress from other dislocations and its image forces, which we obtain in analytical form [47], [48]. Dislocations could only glide on his slip plane and are blocked by the grain boundary.

We did not take into account screw dislocations in the modeling despite the fact that molybdenum has a bcc lattice, since it was shown in [49,50] that edge dislocations in molybdenum have a higher velocity than screw components and therefore have a greater effect on the deformation and fracture processes. Tanaka in his work [51] on the fracture of iron using the 2D dynamics method also models the plastic zone only by edge dislocations. There are eleven dislocation sources in the model, ten in the grain in grain boundary and the other one near the crack tip, fig 7. Since grain boundary sources are often observed in bcc metals [52].

If the stress on the dislocation source exceeded the start stress (friction lattice stress), then two dislocations were emitted from the dislocation source near the crack tip and only one if the dislocation source got locked in the grain boundary. The grain size is defined by the distance between the crack tip and the grain boundary. The grains were assumed to be infinitely long in the *y*-direction.

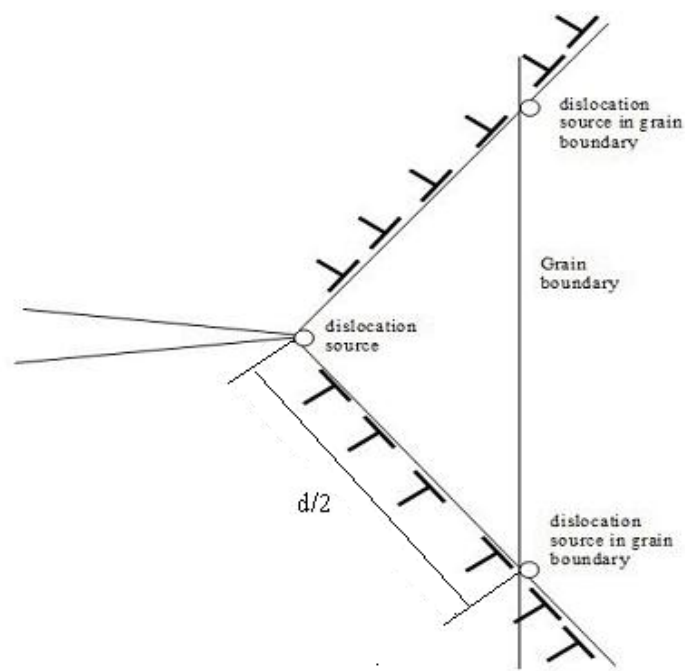

Fig. 7    Plastic zone near the crack tip.

The crack is subjected to pure mode I loading $K_I$ so the stress acting on dislocations from crack tip was calculated according to [30].

$$\sigma_{11} = \sigma_v \sqrt{\frac{a}{2\pi r}}(\cos(\theta/2)(1-\sin(\theta/2)\sin(3\theta/2)))$$

$$\sigma_{22} = \sigma_v \sqrt{\frac{a}{2\pi r}}(\cos(\theta/2)(1+\sin(\theta/2)\sin(3\theta/2))) \qquad (3)$$

$$\sigma_{12} = \sigma_v \sqrt{\frac{a}{2\pi r}}\cos(\theta/2)\sin(\theta/2)\cos(3\theta/2)$$

where $\sigma_v$ is the external stress, $\theta$ – angle between the dislocation position and the x-axis.

The stress on each dislocation was calculated as well as the corresponding dislocation speeds [53]:

$$v = A\sigma_{12},$$

$$A = \frac{2bD_s}{kT}\exp(-F_k/kT) = 2{,}6*10^{-10} m/Pa*c \qquad (4)$$

$D_s$ - is self-diffusion coefficient, $b$ – Burgers vector and $\sigma_{12}$ – stress acting on the dislocation, k – Boltzmann constant and T – temperature. The exact value of A is not very important since it only changes the time step.

The time step was chosen so that the dislocation under maximal stress would pass the distance 100$b$ during this time. Then new displacements of each dislocation were obtained from velocity and time step.

Since the results of the computer modeling [36] and fractography (Fig. 5) show that crack tip stays sharp under loading at average homology temperatures, so the actual value of the stress intensity factor was calculated as in [43]:

$$K_{in} = \sigma_v\sqrt{\pi a} - \frac{3Gb}{2(1-v)\sqrt{2\pi}}\sum_n \frac{\cos(\theta/2)\sin(\theta\theta)}{\sqrt{r}} \qquad (5)$$

where $\sigma_v$ ,-external stress, $n$ is the number of dislocations in the grain containing the crack tip, G – shear modulus,, a –the half-length of the crack.

In our previous manuscript we studied the evolution of the dislocation ensemble near the semi-infinite crack tip in the case when a plastic zone exists only in one grain and external stress is constant [54]. We also showed that, due to dislocations near the crack tip, the stress intensity factor was decreased, and therefore its value depends on the location of the dislocations at that moment, fig.8 [55]. The first dislocation emission occurs at 20% of the critical stress intensity. The same result was obtained by Zeng and Hartmaier [27]. The stress intensity factor increases when a dislocation is emitted, because dislocations are shielding the crack tip from the external stress. In the presence of a nonpenetrable boundary before the dislocation ensemble, the stress intensity factor grows faster [56].

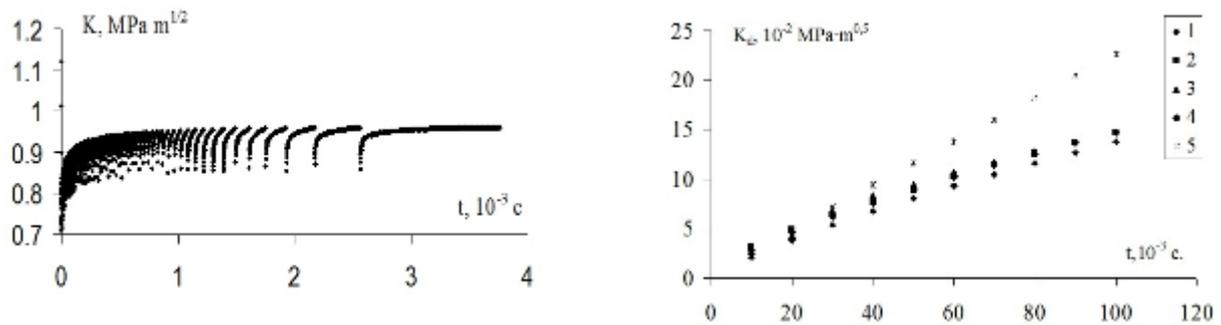

Stress intensively factor via time

Stress intensively factor via different configurations of dislocation ensemble near crack: 1 – no dislocations, 2 – chaotic dislocations, 3 – dislocation wall, 4 – penetrable dislocation wall, nonpenetrable dislocation wall.

Fig. 8    Stress intensity factor vs time.

In our case external stress increased linearly with time until that the material is broken($K_{in} > K_{1c}$), then a critical stress intensity factor was calculate:

$$K_{crit} = \sigma_v \sqrt{\pi a} \qquad (6)$$

The modeling object was molybdenum: $b = 3 \cdot 10^{-10}$ m, $G = 140$ GPa, $K_{1c}=2$ MPa·m$^{1/2}$, friction lattice stress was taken to be 20 MPa at room temperature and

200 MPa for the liquid nitrogen temperature, the size of dislocation sources was 50*b*, crack length 0.2 cm. Load rate was 100 MPa/s.

## 5. Computer modeling results and discussion

Modeling predicts the dependence of the fracture toughness on the grain size (Fig. 9) in good agreement with experimental data that is shown in Fig. 1. At liquid nitrogen temperature the fracture toughness is practically constant, 5,7 MPa·m$^{1/2}$, because in this case the plastic zone is smaller than 10 μm. The number of the dislocation pairs in this case is 236, fig.10.

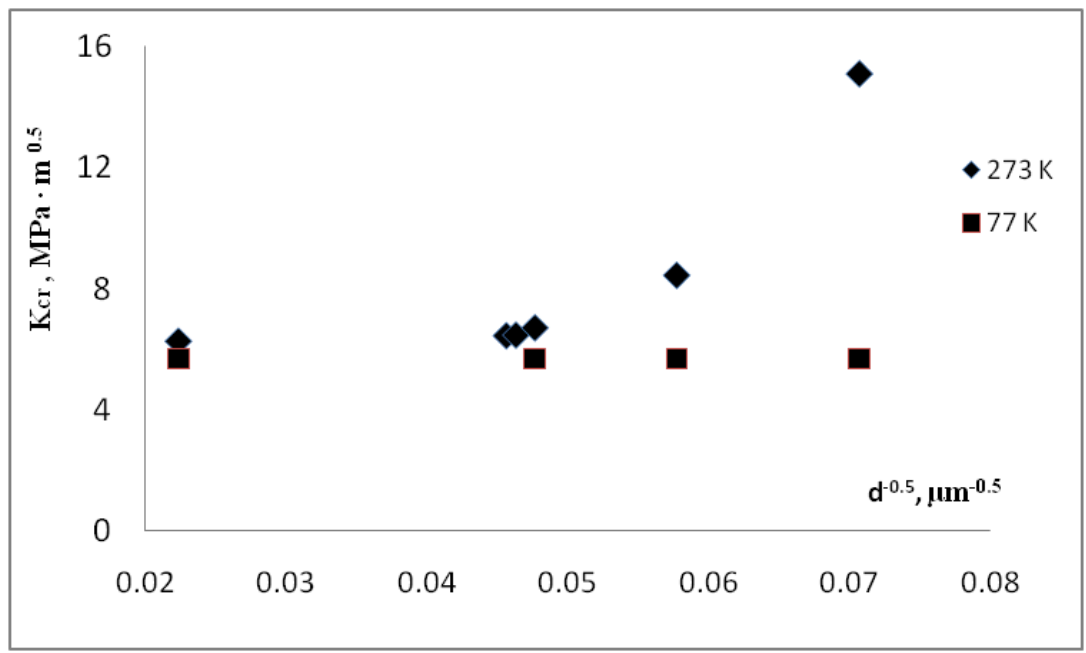

Fig. 9   Dependence of the fracture toughness on the grain size.

At room temperature and large grain size, the fracture toughness is somewhat higher than at 77 K, because at lower values of friction stress, 20 MPa, the size of the plastic zone is 237 μm and the number of dislocations in the plastic zone is around 403.

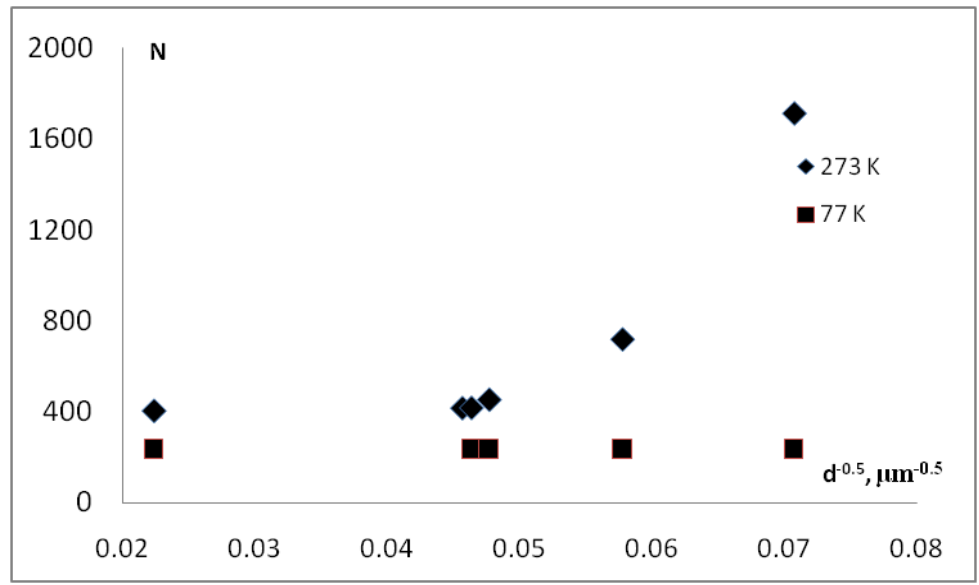

Fig.10  The number of dislocations in the plastic zone.

At the grain size of 220 μm the dislocation source at the grain boundaries starts contributing, however they still emit a small number of dislocations, about 30 pairs, and therefore the fracture toughness remains practically the same. At the grain size of 150 μm the total number of dislocations in the first grain is 461 and in all other grains is 256 pairs, so the fracture toughness begins to grow rapidly because the number of dislocations near the crack tip increases. At the grain size of 100 μm the total number of dislocations is 1710 pairs and size of plastic zone is 507 μm, fig. 10.

The analysis of the distribution of dislocations in the plastic zone (Fig. 11) allowed us to give a physical interpretation of the dependence of crack resistance on grain size. Figure 11 shows the curves normalized so that curve 1 is normalized to unity. In the case of large grain sizes, d3, the normalized dislocation density gradually decreases with distance from the crack head (Fig. 11, curve 1), which is due to the influence of the attractive forces of the image, as it decreases with distance from the vertex. Similar distributions have been observed experimentally [42]., and a similar distribution was obtained by Zeng and Hartmeier [27] using computer modeling. In this case, the screening effect of the crack head dislocations is small and the increase in crack resistance is insignificant. For smaller grain sizes, d2, a cluster is formed near the boundary (Fig. 11, curve 2). But even in the presence of a significant accumulation of dislocations near the boundary, the crack resistance increases by only a few percent, which is due to the fact that the only dislocation source near the crack tip is blocked quite quickly.

At even smaller grain size d1, under the action of the accumulation, the Frank-Reed source in the neighboring grain is triggered and a dislocation pile-up is formed in it. It is under its action that the dislocations are redistributed in the plastic zone of the first grain (Fig. 11, curve 3). The dislocations move to the crack head, block its propagation and significantly increase the crack resistance (in the case of molybdenum, more than twice).

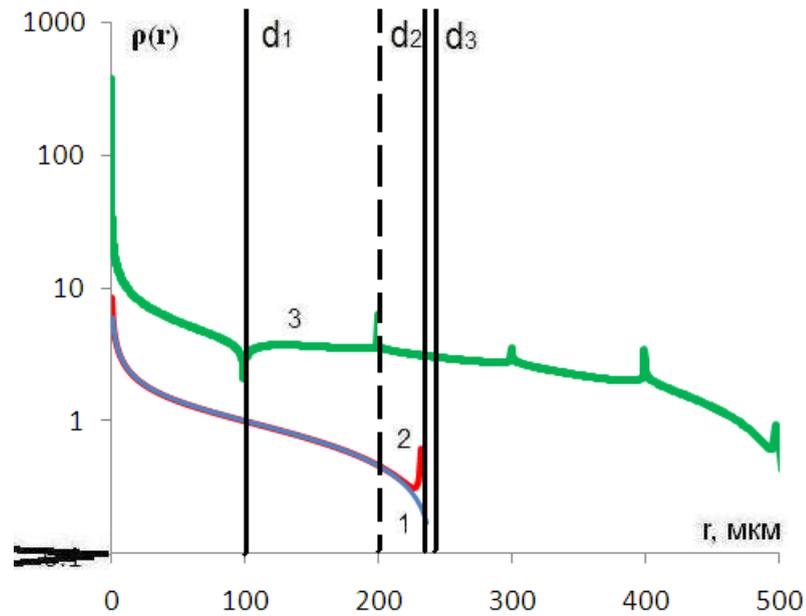

Fig. 11  Dislocations density at different grain sizes.

Thus, the results of our modeling confirm the above-mentioned suggestion that the non-monotonous character of the fracture toughness structure dependence in molybdenum is caused by the interaction of dislocations in the plastic zone with the grain boundary.

### 6. Conclusions:

It is shown that the fracture toughness of molybdenum depends on the grain size at brittle-ductile transition. At the grain size of about 280 μm, fracture toughness abruptly grows from 6,5 to 15 MPa·m$^{1/2}$. By utilizing computer modeling we confirm that the effect is caused by the interaction of the dislocations near the crack tip with the grain boundaries. If the plastic zone touches the grain boundary, the dislocation sources on it start contributing, which results in extension of the plastic zone onto several grains and repulsion of some of the dislocations towards the crack tip, which, in turn, cause almost twofold increase of the fracture toughness.